\documentclass[twocolumn,aps,pra,showpacs,groupedaddress,superscriptaddress,nofootinbib]{revtex4-1}

\usepackage[utf8]{inputenc}  
\usepackage[T1]{fontenc}

\usepackage[normalem]{ulem}

\usepackage[dvipsnames]{xcolor}
\usepackage{graphicx,epsfig}
\usepackage{braket}
\usepackage{color}
\usepackage{amsmath}
\usepackage{amsfonts}
\usepackage{fancyhdr}
\usepackage{mathrsfs}
\usepackage[makeroom]{cancel}
\usepackage{mathtools}
\usepackage{MnSymbol}

\usepackage{subfigure}
\usepackage{float}

\usepackage[pdftex,linkcolor=blue,citecolor=blue,urlcolor=blue,colorlinks]{hyperref}

\newcommand{\mean}[1]{\langle#1\rangle}

\DeclarePairedDelimiter\abs{\lvert}{\rvert}%
\DeclarePairedDelimiter\norm{\lVert}{\rVert}%
\makeatletter
\let\oldabs\abs
\def\abs{\@ifstar{\oldabs}{\oldabs*}}
\let\oldnorm\norm
\def\norm{\@ifstar{\oldnorm}{\oldnorm*}}
\makeatother

\newcommand{\dgr}{^{\dagger}}

\newcommand{\enum}[1]{\mathit{e}^{#1}}

\newcommand{\Spvek}[2][r]{%
  \gdef\@VORNE{1}
  \left(\hskip-\arraycolsep%
    \begin{array}{#1}\vekSp@lten{#2}\end{array}%
  \hskip-\arraycolsep\right)}
\newcommand{\appropto}{\mathrel{\vcenter{
\offinterlineskip\halign{\hfil$##$\cr
  \propto\cr\noalign{\kern2pt}\sim\cr\noalign{\kern-2pt}}}}}

\newcommand{\overbar}[1]{\mkern 1.5mu\overline{\mkern-1.5mu#1\mkern-1.5mu}\mkern 1.5mu}

\begin{document}
\title{Effective triangular ladders with staggered flux from spin-orbit coupling in 1D optical lattices  }
\author{J. Cabedo}
\affiliation{Departament de F\'isica, Universitat Aut\`onoma de Barcelona, E-08193 Bellaterra, Spain.}
\author{J. Claramunt}
\affiliation{Departament de Matem\`atiques, Universitat Aut\`onoma de Barcelona, E-08193 Bellaterra, Spain.}
\affiliation{Departamento de Matem\'atica, Universidade Federal de Santa Catarina, 88040-900 Florian\'opolis SC, Brazil.}
\author{J. Mompart}
\affiliation{Departament de F\'isica, Universitat Aut\`onoma de Barcelona, E-08193 Bellaterra, Spain.}
\author{V. Ahufinger}
\affiliation{Departament de F\'isica, Universitat Aut\`onoma de Barcelona, E-08193 Bellaterra, Spain.}
\author{A. Celi}
\affiliation{Departament de F\'isica, Universitat Aut\`onoma de Barcelona, E-08193 Bellaterra, Spain.}


\begin{abstract}
Light-induced spin-orbit coupling is a flexible tool to study quantum magnetism with ultracold atoms. In this work we show that spin-orbit coupled Bose gases in a one-dimensional optical lattice can be mapped into a two-leg triangular ladder with staggered flux following a lowest-band truncation of the Hamiltonian. The effective flux and the ratio of the tunneling strengths can be independently adjusted to a wide range of values. We identify a certain regime of parameters where a hard-core boson approximation holds and the system realizes a frustrated triangular spin ladder with tunable flux. We study the properties of the effective spin Hamiltonian using the density-matrix renormalization-group method and determine the phase diagram at half-filling. It displays two phases: a uniform superfluid and a bond-ordered insulator. The latter can be stabilized only for low Raman detuning. Finally, we provide experimentally feasible trajectories across the parameter space of the SOC system that cross the predicted phase transition.    
\end{abstract}
\maketitle


\section{Introduction}\label{sec-intro}


Spin-orbit coupled Bose and Fermi gases both in the bulk or loaded in optical lattices are a flexible playground for 
studying many-body physics and quantum phase transitions in a controlled manner.
By entangling internal and external degrees of freedom the spin-orbit coupling (SOC) produced by Raman beams \cite{Lin-2009,Lin-2011} leads already at the single-particle or at the mean-field levels to spatially-dependent dressed states with modified dispersion relation and spatially dependent interactions \cite{Higbie-2002}. 
Such behavior can be interpreted in terms of a synthetic gauge field \cite{Dalibard-2011, Goldman-2014} that can be also density dependent \cite{Edmonds-2013}.

The successful experimental demonstrations of the last decade of synthetic one-dimensional (1D) and two-dimensional (2D) SOC \cite{Zhang-2012, Ji-2015, Wu-2016, Sun-2016} have opened interesting perspectives.  
In the bulk SOC can stabilize exotic phases like the stripe phase \cite{Li-2012,Li-2017}, where translation invariance is spontaneously broken \cite{Li-2015}, 
in analogy with supersolids very recently realized in dipolar gases \cite{Tanzi-2019,Chomaz-2019, Bottcher-2019} (see also \cite{Leonard-2017} for the realization of supersolid-like state in a cavity). Under suitable conditions SOC gives also access to beyond-mean-field dynamics in weakly interacting dilute gases \cite{Cabedo-2019} (for experimental demonstrations of beyond-mean-field effects due to competing interactions see \cite{Cabrera-2017,Cheiney-18,Semeghini-2018}) by inducing spin changing collisions in $F=1$ BEC (see \cite{Williams-2012}) from SU(3)-symmetric density-density interactions that are controllable.   
Interestingly, SOC combined with radio-frequency dressing offers a novel mechanism to achieve subwavelength optical lattices \cite{Anderson-2020}.

Combining SOC and optical lattices makes easier to access the strongly coupled regimes, and more evident the connection to quantum magnetism.
On the one hand, the lattice quenches the kinetic term and allows for (relatively) large interactions and Rabi couplings with negligible losses \cite{Maciej-Book-2012}.
On the other hand, the lattice introduces another length scale $1/k_L$ ($k_L$ is the lattice beam wavevector) in addition to the inverse of the Raman momentum kick $1/k_R$ 
and thus favors frustration.
Such effect becomes evident when the atomic spin states are interpreted as sites of a synthetic dimension \cite{Boada-12,Ozawa-19}.
Raman coupled spin states in 1D spin independent optical lattices experience a synthetic magnetic flux proportional to $k_R/k_L$ that leads to
the appearance of edge states in narrow Hofstadter slabs \cite{Celi-2014}, as experimentally demonstrated in \cite{Stuhl-15,Mancini-15,Livi-16,Kolkowitz-17} 
(also with atomic momentum states \cite{An-17} and with photons \cite{Lustig-19}). 
Remarkably, the correspondence between the Hofstadter model \cite{Hofstadter-76} in 2D lattices and quasi-1D systems extends also to its topological properties and quantum Hall response \cite{Mugel-17} as experimentally demonstrated in \cite{Genkina-19}. Even more striking, such correspondence extends under proper conditions also when interactions are included \cite{Cornfeld-15,Calvanese-19}.

Bosonic flux ladders are the simplest quasi-1D systems that realize such correspondence. 
Already at the single particle level, they provide a toy model of type-II superconductors \cite{Orignac-2001, Granato-2005, Tokuno-2014} and display Meissner and Vortex phases (the latter being the lattice version of the stripe phase) as first experimentally demonstrated  in real-space ladders in \cite{Atala-14}. The interplay between the magnetic flux, the rung vs leg tunnelings, and interactions in real and synthetic ladders leads to a variety of interesting phases and have been extensively studied especially for strong interactions, see for instance \cite{Petrescu-13,DiDio-15,Piraud-2015,Petrescu-2015,Greschner-2016,Piraud-2016,Calvanese-2017,Petrescu-2017,Citro-2018,Greschner-2018,Greschner-prl-2019}. For similar studies in fermionic ladders see for instance \cite{Barbarino-15,Barbarino-16,Junemann-2017,Ghosh-17,Haller-18}.


In this work we consider synthetic flux ladders formed by Raman-dressed spin-1/2 atoms in 1D optical lattices from a different perspective. 
We show that, when the lowest-band approximation holds, they can be mapped into a two-leg triangular ladder with staggered flux. For alternative theoretical proposals of synthetic triangular and zigzag lattices see \cite{Anisimovas-2016-pra,Suszalski-16-pra}, for an experimental realization in synthetic lattices in momentum space with constant fluxes  see \cite{An-18}. The parameters of the effective model, namely, the rung and longitudinal tunnelings and the strength of the flux, can be widely adjusted by tuning the laser dressing parameters. Irrespective of the interactions between the spin states, the interactions in the effective ladder can be onsite. In the experimentally accessible regime of large separation between the bandwidth of the lower band and the gap to the higher band of the original square ladder, we can access both the weakly and the strongly interacting regimes of the triangular ladder within the validity of the mapping. 

Lattices and ladders with triangular geometry have been widely studied in ultracold atoms \cite{Becker-2010} especially in connection to supersolidity 
\cite{Melko-2005, Wessel-2005} and frustrated quantum magnetism \cite{Lacroix-2011}. 
In triangular translational-invariant configurations, the presence of complex tunnelings naturally gives rise to staggered fluxes, for instance equal to $\pi$. 
In optical lattices they can be generated, e.g., by accelerating the lattice potential along a closed orbit, and employed to study classical magnetism in experiments with ultracold bosons \cite{Struck-2013}. In the presence of strong interactions, such systems offer a promising route towards the realization of quantum spin liquid phases \cite{Eckardt-2010-epl}, both in homogeneous gases \cite{Hauke-2010-njp} and in the presence of an harmonic trapping \cite{Celi-2016-prb}. Fully frustrated, i.e., $\pi$-flux, triangular ladders have been theoretically studied at strong coupling in \cite{Mishra-2013,Mishra-2014,Greschner-prb-2019}. 
In the hardcore-boson limit at half filling the corresponding antiferromagnetic $XX$-spin model displays a superfluid phase and a bond-ordered gapped phase 
separated by Berezinskii–Kosterlitz–Thouless-(BKT-)type phase transition \cite{Tonegawa-1990-ptps,Tonegawa-1992-JPSJ}. The latter phase is an analogue of the dimer phase in the Majumdar-Ghosh (MG) model \cite{Majumdar-1969}.  

Here we show that this interesting phase survives also for fluxes close but different from $\pi$ and that it is experimentally accessible in SOC experiments through the mapping to the synthetic flux ladder we introduce. In the latter, the bond-ordered phase appears for (intermediate) interactions, Raman couplings, and detunings that do not seem to have been considered previously in the literature. 




The paper is organized as follows. In Sec. \ref{sec-synthdim}, we discuss the lowest-band approximation in SOC-coupled semisynthetic lattices for 
spin-$S$ bosons by describing its lowest band in terms of an effective 1D lattice model for a quasi particle with complex tunneling terms of various range. In Sec. \ref{sec-spin1/2}, we specialize in the $S=1/2$ case and show that, for a wide regime of parameters, only first- and second-neighbor tunnelings become relevant, that is the rung and leg tunnelings of a triangular ladder. We detail the mapping and its range of validity and show that with very good approximation interactions are local in the  triangular ladder. In Sec. \ref{sec-HCB}, we study the phase diagram of the effective triangular ladder in the hard-core-boson limit and  discuss its bond-ordered phase and its experimental accessibility.
Finally, we summarize our results and comment about future developments in Sec. \ref{sec-outlook}.

\section{Tunable ladder physics in 1D lattices with spin-orbit coupling}\label{sec-synthdim}

\subsection{Synthetic dimensions in lattices with spin-orbit coupling}\label{subsec-synthdim}

We consider a spin-$F$ spinor gas loaded in a one-dimensional (1D) spin-independent optical lattice, generated with a pair of far-detuned counter-propagating laser beams that intersect with an opening angle $\theta_L$. The lattice is characterized by the laser wavelength $\lambda_L$, which defines the lattice spacing $a = \pi/k_L$, with $k_L = 2\pi \cos(\theta_L)/\lambda_L$ being the recoil momentum. We consider a potential depth $V_0$ sufficiently deep so as to consider the tight-binding approximation, yet shallow enough to avoid the suppression of nearest-neighbour tunneling. That is, we consider $5 E_L < V_0 < 10 E_L$, where $E_L = \hbar^2 k_L^2/2m$ is the recoil energy (here $m$ is the atomic mass).  

One-dimensional Rashba-Dresselhaus (RD) spin-orbit coupling (SOC) is induced in the spinor gas by means of Raman dressing.  Likewise, such dressing is generated by an additional pair of laser beams with wavelength $\lambda_R$ and opening angle $\theta_R$, giving an associated Raman recoil momentum $k_R = 2\pi \cos(\theta_R)/\lambda_R$. Typically, RD SOC can be achieved by coupling $2S+1$ of the $2F+1$ states $\ket{F,m}$ within a given hyperfine manifold of total angular momentum $F$, which are separated via Zeeman splitting, and serve as effective spin states of pseudo-spin $m$. In the absence of inter-atomic interactions, the system can be described by the Hamiltonian \cite{Celi-2014}
\begin{multline}\label{ham0}
H_\mathrm{n.i.} =  \sum_{n,m}\Big( -t\enum{-i\gamma m} a_{n+1,m}\dgr +  \Omega_{m} a_{n,m+1}\dgr \\
+\frac 12 \Delta_m  a_{n,m}\dgr \Big) a_{n,m} +{\rm H.c.},
\end{multline}
where $a_{n,m}\dgr$ and $a_{n,m}$ are, respectively, the creation and annihilation operators for the tight-binding mode at the lattice site $n$ with spin state $m$. Here, $t \sim \frac{4}{\pi}E_L\left( V_0/E_L\right)^{3/4}\mathrm{e}^{-2(V_0/E_L)}$ \cite{Bloch-review-2008} is the tunneling rate between the nearest-neighbor (NN) modes, $\Omega_m$ is the Raman coupling strength between levels $m$ and $m+1$, $\Delta_m$ is an onsite energy shift that depends on the detuning of the Raman lasers, and $\gamma = 2k_R a$. The strength of the Raman dressing is constrained by the tight-binding consideration. To be consistent, the coupling strengths $\Omega_m$ are required to be much smaller than the energy splitting between the tightly-bound states and the rest of single-particle states, roughly given by $\epsilon_\mathrm{t.b.} \sim 2\sqrt{V_0 E_L}$ .

As noted in \cite{Celi-2014}, such system is equivalent to a 2D lattice where the internal spin states act as a synthetic dimension, pierced by an effective magnetic field with flux $\phi = \gamma/2\pi$. That is, it realizes the Hofstadter model on a slab and reproduces the main features of magnetic lattice systems, such as the fractal Hofstadter-butterfly spectrum and the chiral edge states of the associated Chern insulating phases. Interactions in synthetic dimensions have naturally a long-range character as particles with different spins interact locally when they occupy the same site in the actual 1D lattice. In fact such long-range behavior can be alterated or even suppressed, e.g., displacing spatially the spin states as originally proposed in \cite{Boada-12}, or considering non-SU($F$) interactions for the spin states as obtained from Feshbach resonance (see for instance \cite{Cabrera-2017}) or by properly modulating the scattering length as recently proposed in \cite{Barbiero-2019}.

By contrast, here we show that, following a truncation of the single-particle Hilbert space, interesting quasi 1D structures can be obtained where the interactions and tunnelings can be controlled by adjusting the dressing parameters. In our model the synthetic gauge field induced by the Raman kick plays a crucial role. Dressing with radio-frequency (rf)-couplings that do not transfer momentum can be still employed to tune the scattering properties of the gas, as recently experimentally demonstrated with potassium atoms in \cite{Sanz-2019}.

\subsection{Effective quasi-particle from Raman dressing}\label{subsec-lowestband}

\begin{figure*}[t]
\includegraphics[width=0.95\linewidth]{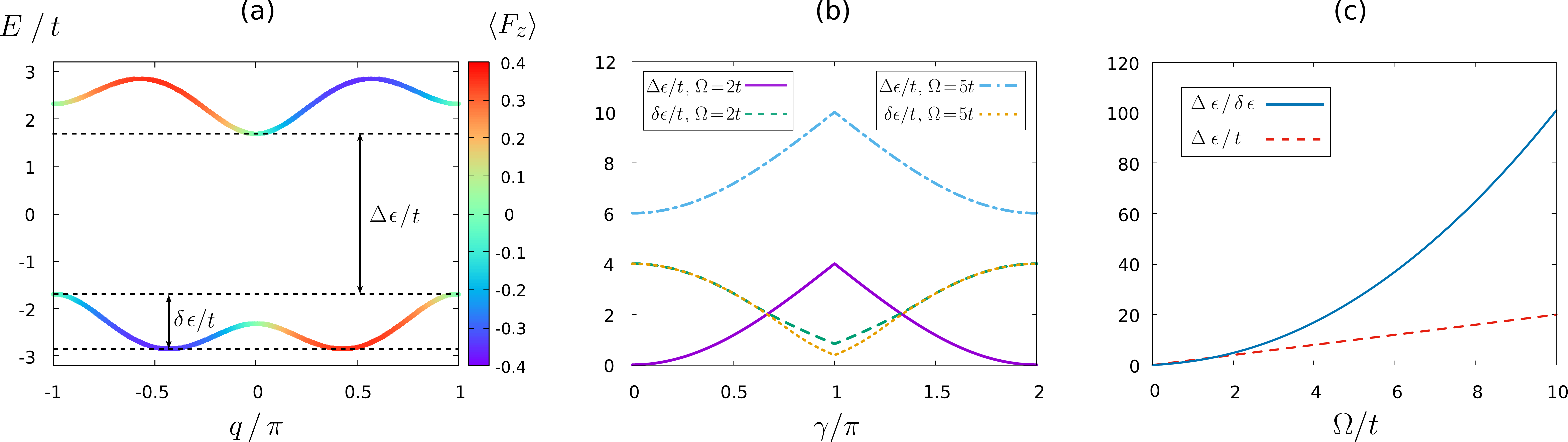}
\caption{(Color online) \textbf{Noninteracting energy scales:} (a) Energy bands for the case $S = 1/2$ with $\Omega = 2t$, $\gamma=0.9\pi$, and $\delta=0$. The arrows indicate the bandwidth $\delta\epsilon$, which is related to the tunneling strength of the effective quasiparticle, and the band-gap $\Delta \epsilon$ that separates the two bands. The lowest band truncation requires that $\Delta\epsilon \gg U$. The color texture represents the expected value of the spin of the band states. (b) $\Delta \epsilon$ and $\delta \epsilon$ as a function of $\gamma$ for $\Omega = 2t$ and $\Omega = 5t$. The ratio $\Delta\epsilon/\delta\epsilon$ is maximized at $\gamma=\pi$ and increases with $\Omega$. (c) $\Delta \epsilon$ and $\Delta \epsilon /\delta \epsilon$ as a function of $\Omega$ at $\gamma = \pi$. In all figures $\Delta_m = 0$.}\label{bandscales} 
\end{figure*}

Hamiltonian \eqref{ham0} is block diagonal in orthogonal quasimomentum  subspaces
\begin{equation}\label{ham1}
H_\mathrm{n.i.} = \sum_q H_q,
\end{equation}
with
\begin{align}\label{hamq}
H_q = & \sum_{m} \big(-2t \cos{(q + \gamma m)} + \Delta_m \big) \tilde{a}_{q,m}\dgr \tilde{a}_{q,m} + \cr
& + \sum_m \left( \Omega_{m-1} \tilde{a}_{q,m-1}\dgr \tilde{a}_{q,m} + \Omega_{m-1}^* \tilde{a}_{q,m}\dgr \tilde{a}_{q,m-1} \right), \cr
\end{align}
where we have introduced the Fourier transformed modes $\tilde{a}_{q,m}\dgr = \frac{1}{\sqrt{L}}\sum_n \enum{iqn} a_{n,m}\dgr$, with $L$ being the total number of sites in the lattice. Hamiltonian \eqref{ham1} has $2S+1$ bands, which we label as $\epsilon_{q,m'}$, with $m' \in \left\{0,1,...,2S \right\}$ ($\epsilon_i \leq \epsilon_j$ for $i < j$), and with associated band modes
\begin{equation}\label{bandmodes}
\tilde{b}_{q,m'}\dgr = \sum_{m} \text{U}_{m',m}(q) \tilde{a}_{q,m}\dgr .   
\end{equation}

Here, $$\text{U}(q) = \sum_{m_1,m_2} \text{U}_{m_1+F,m_2}(q) \big( \tilde{a}_{q,m_2}\dgr \ket{0} \big) \big(\bra{0}\tilde{a}_{q,m_1} \big)$$ is the unitary transformation that relates the dressed eigenbasis $\{ \tilde{b}_{q,m'}\dgr \ket{0} \}_{m'}$ with the uncoupled hyperfine state basis $\{ \tilde{a}_{q,m}\dgr \ket{0} \}_m$, with $\text{U}_{m',m}(q) = \bra {0}\tilde{a}_{q,m} \tilde{b}_{q,m'}\dgr\ket{0}$. Without loss of generality, we can assume the coefficients $\text{U}_{ij}(q)$ to be real. 

We now restrict ourselves to the regime where the lowest band can be well separated from the higher energy bands. This occurs for sufficiently large coupling coefficients $\Omega_m$, with a band gap that depends also on the value of the phase $\gamma$. Under these circumstances, the low-energy landscape of the system is well described by the truncated Hamiltonian
\begin{equation}\label{hamqtruncated}
H_\mathrm{n.i.} \simeq \sum_q \epsilon_{q,0}\tilde{b}_{q,0}\dgr\tilde{b}_{q,0}.
\end{equation}

We now introduce the inverse Fourier-transformed truncated basis
\begin{equation}\label{ftbandmodes}
b_n\dgr := \frac{1}{\sqrt{L}}\sum_q\enum{-iqn}\tilde{b}_{q,0}\dgr.
\end{equation}
Substituting \eqref{ftbandmodes} into \eqref{hamqtruncated} yields
\begin{equation}\label{hamqtruncatedx}
H_\mathrm{n.i.} \simeq -\sum_n \sum_l t_{l} b_{n+l}\dgr b_{n},
\end{equation}
with
\begin{equation}\label{tunnelingcoeff}
t_{l} = -\frac{1}{L}\sum_q\enum{-iql}\epsilon_{q,0}.
\end{equation}
The effective dispersion band $\epsilon_{q,0}$ has a shape that depends on the total spin size $S$ and that can be tailored by adjusting the parameters $\Omega_m$, $\gamma$ and $\Delta_m$. Thus, the strength and relative phase of the different tunneling coefficients in the truncated basis, $t_{l}$, can be manipulated.

Likewise, as long as the energy per particle is much smaller than the gap between the two bands, the interaction Hamiltonian can be re-expressed in the truncated basis $\left\{b_{n}\dgr\ket{0}\right\}_n$. For simplicity we will assume SU$(F)$ symmetric interactions, which is a good approximation, for instance, for $F=1$ $^{87}$Rb. In the lattice, the tight-binding interaction Hamiltonian reads
\begin{equation}\label{intham0}
H_\mathrm{int} = \frac{U}{2}\sum_n N_n(N_n - 1) = \frac{U}{2}\left(\sum_n N_n^2 - N\right), 
\end{equation}
with
\begin{equation}\label{densityoperator0}
N_n = \sum_{m} a_{n,m}\dgr a_{n,m},
\end{equation}
where $U$ is the onsite interaction energy per particle pair and $N$ is the total number of particles.

From \eqref{bandmodes} and \eqref{ftbandmodes} it follows that
\begin{align}
a_{n,m}\dgr &= \frac{1}{\sqrt{L}} \sum_q \enum{-iqn} \tilde{a}_{q,m}\dgr \simeq \frac{1}{\sqrt{L}} \sum_q \enum{-iqn} \text{U}_{0,m}(q) \tilde{b}_{q,0}\dgr \cr
&= \frac{1}{L} \sum_l \left( \sum_q \enum{iq(l-n)} \text{U}_{0,m}(q) \right) b_l \dgr = \sum_l \lambda_m^{(l)} b_{n+l}\dgr.
\end{align}

In the last equality we have defined the coefficients
\begin{equation}
\lambda_m^{(l)} := \frac{1}{L}\sum_q \enum{iql} \text{U}_{0,m}(q),
\end{equation}
which correspond to the amplitudes of the modes $b_n\dgr\ket{0}$ at sites $n+l$. Therefore, we can define the coefficients $C_{l,l'} = \sum_m \lambda_m^{(l)} ( \lambda_m^{(l')})^*$ and rewrite $N_n$ as
\begin{equation}\label{densityoperator1}
N_n = \sum_{l,l'} C_{l,l'} b_{n+l}\dgr b_{n+l'}.
\end{equation}

For weakly coupled gases, the spread of the truncated modes can be significant, while they become tightly localized in the strong coupling limit. Thus, at weak couplings the density-density terms in the interaction Hamiltonian \eqref{intham0} may include significant higher order terms in the truncated basis. In practice, however, the lowest-band truncation demands relatively large $\Omega_m$, for which we can safely truncate the total Hamiltonian to 
\begin{equation}\label{hamtotalspinF}
H \simeq \sum_n \sum_l \left(-t_{l} b_{n+l}\dgr b_{n} + \frac{U C_{0,0}}{2}\sum_{l'}C_{l,l'}b_{n}\dgr b_{n}b_{n+l}\dgr b_{n+l'}\right).
\end{equation}
In this way, following a lowest-band truncation, the Hamiltonian for a SOC gas in a lattice can be interpreted as an effective Hubbard model that may include long-range complex tunneling terms and non-trivial interactions. In the next sections, we explore the simplest of these systems, the spin-1/2 case. In Sec. \ref{sec-spin1/2}, we show that the effective Hamiltonian \eqref{hamtotalspinF} then describes a triangular ladder with tunable staggered flux. In Sec. \ref{sec-HCB}, we study its phase diagram and show that through the mapping we can identify a non-trivial many-body phase of the original spin-1/2 flux ladder with interactions, a bond-ordered phase, that was unnoticed.

\begin{figure*}[t]
\includegraphics[width=0.95\linewidth]{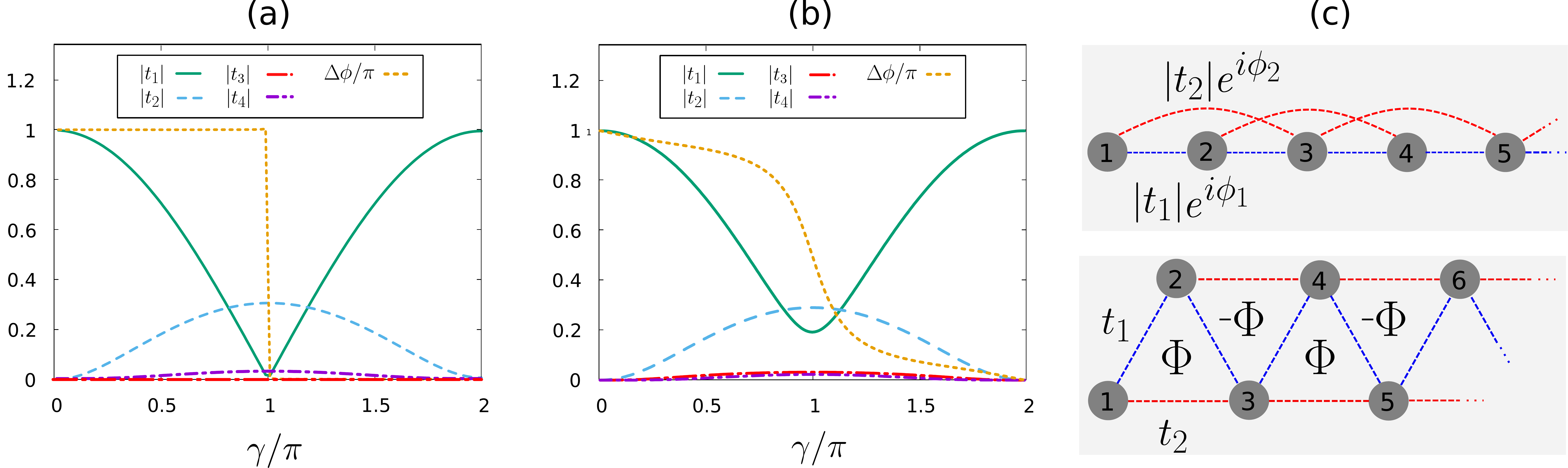}
\caption{(Color online) Effective triangular ladder with staggered flux:} (a - b) Largest contributions to the non-interacting Hamiltonian \eqref{hamqtruncatedx} for $S=1/2$. In solid green, dashed blue, long-dashed-dotted red, and dashed-dotted purple are plotted, respectively, the absolute value of the tunneling coefficients $t_1$, $t_2$, $t_3$, and $t_4$, as a function of the Raman flux $\gamma$ for $\Omega = t$.  In dotted yellow, it is plotted the relative phase between $t_2$ and $t_1$, $\Delta\phi$. The detuning is set to $\delta=0$ (a) and $\delta=0.75t$ (b). The tunneling coefficients $t_l$ are, in both cases, scaled to the tunneling energy $t$. (c) When $|t_4|\ll |t_1|, |t_2|$, the system can be mapped into a triangular ladder with staggered flux $\Phi = \phi_2 - 2\phi_1$.\label{triangularladder} 
\end{figure*}

\section{Spin-1/2: triangular ladder with staggered flux}\label{sec-spin1/2}

The spin-$1/2$ can be realized by having only two states of a hyperfine manifold coupled by Raman transitions, with the rest being set off resonance via the quadratic Zeeman shift \cite{Lin-2011}. In this case, $m= \pm 1/2$, and 
\begin{align}\label{hamqspin12}
H_q = &\left(2t\sin\left(\frac{\gamma}{2}\right)\sin(q)+\frac{\delta}{2}\right)\sigma_z \cr & -2t\cos\left(\frac{\gamma}{2}\right)\cos(q) + \Omega\sigma_x,
\end{align}
where $\sigma_i$ are the Pauli matrices and $\delta$ is the detuning, $\Delta_{\pm} = \pm \delta/2$. From \eqref{hamqspin12} it follows that the two bands are given by
\begin{align}\label{lowestband_s12}
\epsilon_{q,\pm} = & \pm \sqrt{\left(2t\sin\left(\frac{\gamma}{2}\right)\sin(q) + \frac{\delta}{2}\right)^2 + \Omega^2} \\ &-2t\cos\left(\frac{\gamma}{2}\right)\cos(q) \nonumber.    
\end{align}

These two bands for the spin-$1/2$ case are represented in Fig. \ref{bandscales}(a) for $\Omega = 2t$, $\gamma = 0.9\pi$ and $\delta=0$. For large enough Raman coupling strength, the bands are separated by a gap that we label as $\Delta\epsilon$ and both have equal intra-band energy width that we label as $\delta\epsilon$. Besides the tight-binding energy $\epsilon_\mathrm{t.b}$, these two quantities define the energy scales of the effective system at the single-particle level. In Fig. \ref{bandscales}(b) and Fig. \ref{bandscales}(c) we plot $\Delta\epsilon$ as a function of $\gamma$ and $\Omega$, respectively. In the vicinity of $\gamma \sim \pi$, the band gap is roughly given by $2\Omega$. In order to fulfil both the tight-binding and the lowest-band truncations, we require $\epsilon_\mathrm{t.b.} \gg \Omega \gg U$. The bandwidth $\delta\epsilon$, on the other hand, is directly related to the strength of the effective tunnelings $t_{l}$ and it is minimized at $\gamma = \pi$, as shown in Fig. \ref{bandscales}(b). The ratio between the band gap and the band width increases fast with $\Omega$ (Fig. \ref{bandscales}(c)).

By substituting \eqref{lowestband_s12} into \eqref{tunnelingcoeff}, we retrieve the effective tunneling coefficients $t_l = \abs{t_l}\enum{i\phi_l}$. In Fig. \ref{triangularladder}, we plot $\abs{t_{l}}$ for $l = 1,2,3$, and $4$, and the relative phase between the coefficients $t_1$ and $t_2$, $\Delta\phi=\phi_2-\phi_1$, as a function of the Raman phase $\gamma$ for $\Omega = t$ and $\delta=0$ (a) and $\delta=0.75 t$ (b). In the regimes that we study here, these four terms are the largest contributions to Hamiltonian \eqref{hamqtruncatedx}, excluding the on-site term proportional to $t_0$. The rest of terms in \eqref{hamqtruncatedx} can be directly neglected as they are orders of magnitude smaller.The terms proportional to $t_4$, which acts as a 4-site range tunneling, can be significant with respect to $t_1$ near $\gamma = \pi$ and small $\delta$ and $\Omega$. Likewise, the coefficient $t_3$, which vanishes at zero detuning, can get comparatively large for moderate values of $\delta$. Still, neither $t_3$ nor $t_4$ become dominant in any range of parameters. Furthermore, in the relative large $\Omega$ regime we are interested in, such terms can be safely neglected (more details in Sec. \ref{sec-HCB}). Then, the non-interacting Hamiltonian is simply given by
\begin{equation}\label{niham_ladder}
H_\mathrm{n.i.} = - \sum_n \big(t_1 b_{n+1} \dgr b_{n} + t_2 b_{n+2} \dgr b_{n} \big) + {\rm H.c.} 
\end{equation}
In this way, the system becomes analogous to a triangular ladder configuration with gauge invariant staggered flux $\pm \Phi = \pm(\phi_2 - 2\phi_1)$, as schematically represented in Fig. \ref{triangularladder}(c). The staggered flux $\abs{\Phi}$ depends on the Raman detuning $\delta$ and $\gamma$, and for $\delta\neq 0$ it ranges from $0$ when $\gamma=(2k+1)\pi$ to $\pi$ when $\gamma=(2k)\pi$, with $k\in \mathbb{Z}$. When $\delta = 0$, $\Phi$ can only take the values $\pm \pi$, which is equivalent to have a uniform flux of $\pi$ across the ladder. In the gauge chosen in the beginning of Sec. \ref{sec-synthdim}, $\phi_2 = \pi$, and $\phi_1 = \frac{\pi - \Phi}{2}$.
\begin{figure*}[t]
\includegraphics[width=0.90\linewidth]{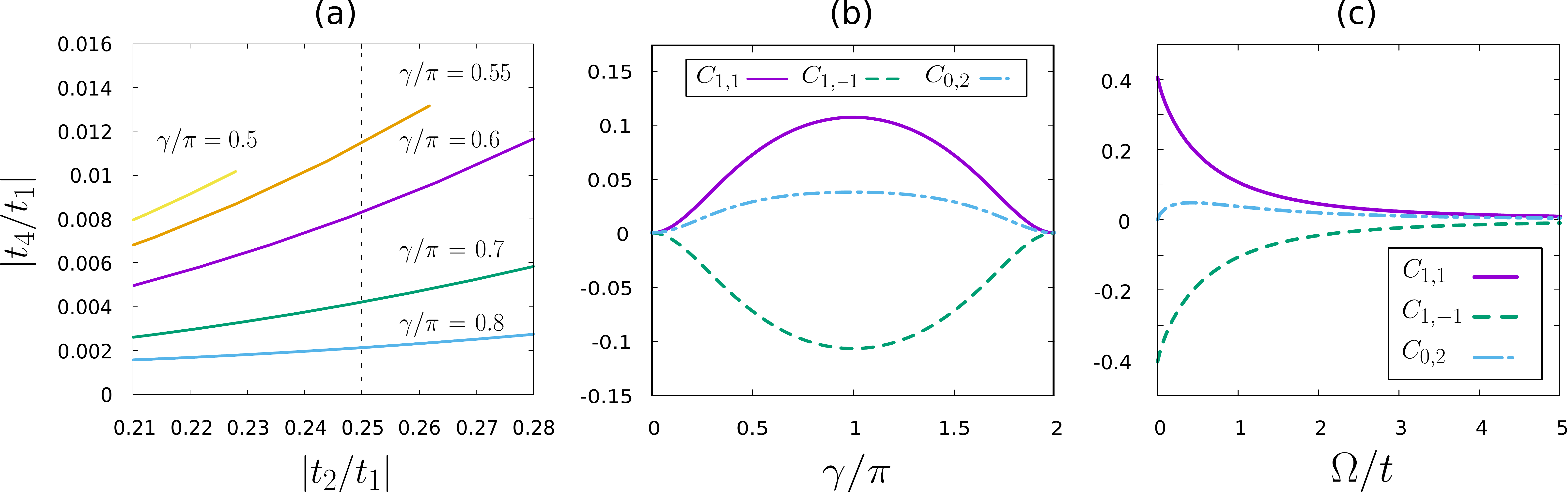}
\caption{(Color online) \textbf{Next-to-leading-order terms in the truncated Hamiltonian:} (a) $|t_4/t_1|$ as a function of $|t_2/t_1|$ at $\delta = 0$ near the single-minimum to two-minima transition at $|t_2/t_1| = 0.25$, for different fixed values of $\gamma/\pi$ $(0.5, 0.55, 0.6, 0.7 , 0.8)$. Only the points in $(\Omega, \gamma)$ that yield a positive band-gap $\Delta\epsilon$ are included. (b) Coefficients $C_{l,l'}$ in the Fourier expansion of the density operator \eqref{densityoperator1} as a function of $\gamma$ for $\Omega = t$, and $\delta=0$. Only the three largest next-to-leading-order coefficients are represented, scaled to the leading term coefficient $C_{0,0}$. (c) $C_{1,1}$, $C_{1,-1}$ and $C_{0,2}$ as a function of $\Omega$ for fixed $\gamma = \pi$ and $\delta=0$, scaled to $C_{0,0}$.}\label{interactcoefficients} 
\end{figure*}

The effective band of the truncated ladder, which we label as $\epsilon'_{q}$, can be written in terms of the new parameters $t_1$ and  $t_2$. In the limit $L \rightarrow \infty$, the band reads
\begin{equation}\label{effectiveband}
\epsilon'_{q} \sim -2\abs{t_1} \cos(q-\phi_1) - 2\abs{t_2} \cos(2q-\phi_2).
\end{equation}
The characteristic Meissner (single band minimum) to Vortex (two band minima) transition for $\delta=0$ ($\Phi = \pm\pi$) in SOC gases \cite{Orignac-2001} occurs for Hamiltonian \eqref{niham_ladder} at $\abs{t_2/t_1} = 1/4$ when $L \rightarrow \infty$. The value is found by solving $\frac{\partial^2 \epsilon'_{q}}{\partial q^2}=0$. Around this value, $\abs{t_4} \ll \abs{t_1},\abs{t_2}$, as shown in Fig. \ref{interactcoefficients}(a) for different trajectories $t_2(\Omega)$ at $\delta=0$ and $\gamma$ fixed. Thus, the transition is accurately captured in the effective triangular ladder truncation. In the figure, only the points fulfilling $\Delta \epsilon > 0$ are included. Observe that when $\gamma$ is made smaller, the trajectories $t_2(\Omega)$ can no longer cross the transition while fulfilling the gap condition.

Likewise, in this situation $\text{U}(q) = \enum{i(\theta_q/2)\sigma_y}$, where $\sigma_y$ is the Pauli matrix and where $\tan(\theta_q) = -2\Omega/(4t\sin(q)\sin(\gamma/2)+\delta)$, and we have
\begin{equation}
\text{U}_{0,+1/2} = \cos(\theta_q/2) \text{ , } \text{U}_{0,-1/2} = -\sin(\theta_q/2).
\end{equation}

From their Fourier series, we obtain the corresponding coefficients $C_{l,l'}$ in \eqref{densityoperator1}, which are plotted in Fig. \ref{interactcoefficients}(b) as a function of $\gamma$ for $\Omega=t$ and scaled to the largest term $C_{0,0}\leq1$. Only the three next largest contributions are included, $C_{1,1}$, $C_{1,-1}$, and $C_{0,2}$, with the rest of them being orders of magnitude smaller. 

When $\Omega \sim t$, the coefficient $C_{0,0}$ is roughly one order of magnitude larger than the next leading terms. Hence, for $\Omega \geq t$ it is safe to keep only the terms proportional to $C_{0,0}$ and rewrite the interacting Hamiltonian for the effective ladder as
\begin{align}\label{ladderhamiltonian_int}
H_\mathrm{int} \simeq &\sum_n \left( U_0\overbar{N}_n\overbar{N}_n + U_1\overbar{N}_{n}\overbar{N}_{n+1} \right) \cr
+&\sum_n \left(-U_1\overbar{N}_n b_{n+1}\dgr b_{n-1} + U_2\overbar{N}_{n} b_{n+2}\dgr b_{n} + {\rm H.c.} \right),\cr
\end{align}
where we drop the contribution proportional to the total number of particles. Here we define the density operator in the truncated basis as $\overbar{N}_n = b_n\dgr b_n$ and the coefficients
\begin{align}
2U_0 &= U C_{0,0}C_{0,0}, \cr
2U_1 &= U C_{0,0}C_{1,1} \simeq -U C_{0,0}C_{1,-1}, \cr
2U_2 &= U C_{0,0}C_{0,2}, 
\end{align}
where we use $C_{l,l'} = C^*_{l',l}$. In Fig. \ref{interactcoefficients}(c) we show the ratio between the different coefficients as a function of $\Omega$. Both $U_1/U_0$ and $U_2/U_0$ decrease fast when $\Omega/t$ is increased, and $C_{0,0}$ approaches $1$. For large $\Omega$, the total effective Hamiltonian can be truncated to
\begin{equation}\label{triangladderhamiltonian}
H \approx \sum_n\left[-\left(t_1 b_{n+1}\dgr b_n + t_2 b_{n+2}\dgr b_n + {\rm H.c.}\right) + \frac{U}{2}\overbar{N}_n^2\right], 
\end{equation}
which is analogous to a triangular ladder configuration with onsite interactions. As shown in Fig. \ref{triangularladder}, depending on the parameters of the system, the ratio $\abs{t_2/t_1}$ can range from $0$ to $+ \infty$. The interaction strength, on the other hand, is limited by the condition on the band gap, roughly $\Delta\epsilon \gg UN_n \sim U \overbar{N}_n$. However, as $\Omega$ is increased, the energy gap increases while the band range $\delta\epsilon$ decreases (see Fig. \ref{bandscales}(c)), and so the maximum allowed interaction strength is increased while all the tunneling coefficients decrease fast. Therefore, both regimes, $U_0 \gg |t_1|, |t_2|$ and $U_0 \ll |t_1|, |t_2|$, can in principle be explored within the lowest band approximation, even if $U \sim t$. It is worth mentioning that, for a given lattice depth $V_0$, the interaction strength can be independently tuned by adjusting the transverse confinement. In the next section we will focus on the strongly interacting regime within the effective ladder, which can be mapped into a frustrated triangular spin ladder with adjustable flux.

\begin{figure*}[t]
\includegraphics[width=0.90\linewidth]{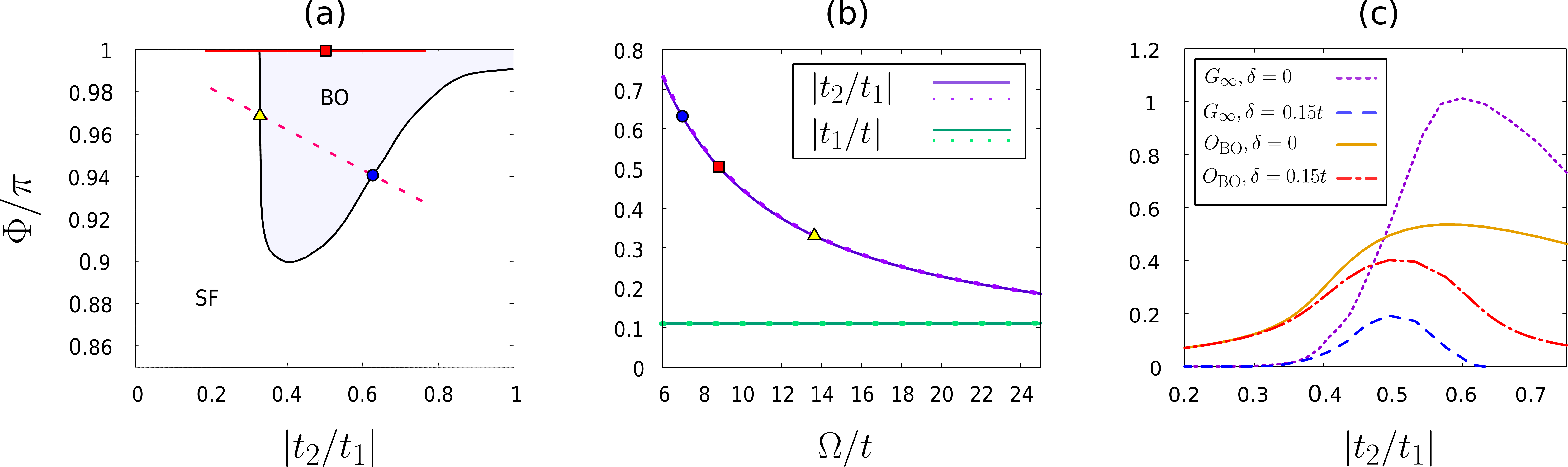}
\caption{(Color online) \textbf{Phase diagram in the hard-core boson limit:} (a) Phase diagram indicating the boundaries between the SF and BO phases in the $|t_2/t_1|$ - $\Phi$ plane. In solid (dashed) red it is plotted a trajectory in parameter space at $\delta = 0$ ($\delta = 0.15 t$), $\gamma/\pi = 0.93$ and $\Omega \in [0,25t]$ for which condition \eqref{numHCBcondition} is fulfilled. The red square indicates the position of the MG point. The yellow triangle and blue circle highlight the position of the boundaries of the BO phase across the $\delta=0.15t$. (b) Effective tunneling ratio $|t_2/t_1|$ (purple lines) and tunneling strength $|t_1|/t$ (green lines) as a function of $\Omega$ and for $\gamma/\pi = 0.93$, in correspondence to the trajectories highlighted in (a), where $\delta = 0$ (solid dark) and $\delta = 0.15 t$ (dotted light). The blue circle, yellow triangle, and red square are in correspondence to figure (a). (c) Extrapolated energy gap in the thermodynamic limit $G_{\infty}$ and bond-order parameter for a chain of size 100, $O_{BO}(100)$ as a function of $|t_2/t_1|$ across the trajectories highlighted in (a): for $\delta=0$ in solid and dotted lines, respectively, and $\delta=0.15t$ in dashed and dashed-dotted lines respectively.}\label{HCB_regime} 
\end{figure*}

\section{Strongly interacting limit: hard-core boson regime}\label{sec-HCB}

For sufficiently large Raman coupling $\Omega$, the lowest-band dispersion is largely suppressed with respect to the gap to the second band (see Fig. \ref{bandscales}(c)), such that $\Delta\epsilon \ggg \delta\epsilon$. In this regime we can consistently consider strong interactions within the effective triangular ladder described by Hamiltonian \eqref{triangladderhamiltonian} and study it in the hard-core-boson (HCB) approximation.  That is, for $\Delta\epsilon\gg U \gg \abs{t_1},\abs{t_2}$  double occupancies in the  triangular ladder are largely suppressed, such that $(b_{i}\dgr, b_j) \rightarrow (S_{i}^+, S_{j}^-)$ and \eqref{triangladderhamiltonian} can be mapped into $J_1-J_2$ $XX$-Hamiltonian on a chain
\begin{align}\label{hardcoreham}
H \sim -\sum_j \left( t_1 S_{j}^{-}S_{j+1}^{+} + t_2 S_{j}^{-}S_{j+2}^{+} + {\rm H.c.} \right).
\end{align}

For $\abs{\Phi} = \pi$, or equivalently, $t_1/t_2 < 0$, such chain has been studied in detail in \cite{Mishra-2013, Mishra-2014}. There, the authors find that the system at half-filling presents two distinct phases in the absence of nearest-neighbor interactions. At $\abs{t_2} \approx \abs{t_1}/3$, the system undergoes a BKT type phase transition from a gapless superfluid (SF) to a gapped bond-ordered (BO) phase. The bond-ordered phase is characterized by a nonzero value in the thermodynamic limit of the bond-order parameter
\begin{equation}\label{BOparam}
O_{BO}(L) = \frac{1}{L}\sum_j (-1)^j \left \langle S_{j}^+S_{j+1}^- + S_{j}^-S_{j+1}^+ \right\rangle .
\end{equation}
Furthermore, the system is solvable at $t_2 = -t_1/2$, where it dimerizes in the presence of frustration and is analogous to the Majumdar-Ghosh (MG) model \cite{Majumdar-1969}. 

Such scenario is reproduced by Hamiltonian \eqref{hardcoreham} when $\delta=0$. Moreover, at nonzero detuning, the presence of a staggered flux modifies the stability of the BO phase and the region for which the SF-BO phase transition occurs. In Fig. \ref{HCB_regime}(a) we plot the SF-BO phase diagram of \eqref{hardcoreham} as a function of $\abs{t_2/t_1}$ and $0<\Phi<\pi$ (we exploit the reflection symmetry of \eqref{hardcoreham}). Similarly as done in \cite{Mishra-2013}, the boundary of the gapped phase is located by computing the single-particle excitation gap
\begin{equation}
G_{L} = E(L,N + 1) + E(L,N-1) - 2E(L,N),
\end{equation}
where $E(L,N)$ is the energy of the ground state for a ladder of $L$ sites and $N$ particles. The onset of a nonzero value of the single-particle excitation energy gap in the thermodynamic limit, $G_{L\rightarrow \infty}$, sets the boundary separating the SF and BO phases. The value of $G_{L\rightarrow \infty}$ is extrapolated by polynomial fitting from the gap of finite sized spin chains. As described in \cite{Mishra-2011-prb}, this allows to compute the gap at low computational cost with DMRG. For the results shown in Fig. \ref{HCB_regime} we compute $G_L$ using the ITensor library for chains with $2L \in [160, 240]$, with $8$ sweeps and bond dimension up to $D=300$ to reach convergence. We consider the sampled points within the gapped phase when the extrapolated value $G_{\infty}>10^{-3}$. The BO phase arises from frustration, and the region in which the BO phase stabilizes rapidly vanishes when $\pi -\Phi$ increases.

In Fig. \ref{HCB_regime}(a), the lines that cross the boundary correspond to two trajectories in parameter space ($\Omega$, $\gamma$, $\delta$) for which the HCB regime can be considered: the onsite interaction energy per pair is constrained to fulfill
\begin{equation}\label{numHCBcondition}
\Delta\epsilon \geq 10^2 \, \mathrm{max}(\abs{t_1}, \abs{t_2}),
\end{equation}
that is, we require that there exist at least an energy separation between the band gap and the effective tunnelings of two orders of magnitude. The solid (dashed) line collects the set of points in parameter space obtained at $\gamma/\pi = 0.93$ and $\delta = 0$ ($\delta = 0.15 t$) by varying the Raman coupling strength from $\Omega=0$ up to $\Omega= 25t$ that fulfill condition \eqref{numHCBcondition}. Larger values of $\Omega$ are disregarded, as they challenge the tight-binding assumption made in Sec. \ref{sec-synthdim} in the regimes considered for $V_0$. Naturally, this is not a fundamental limit, and can be overcome by having a larger lattice depth, $V_0>10$, yet at the cost of scaling down the energy scale of the lattice, $t$. The corresponding values of $\abs{t_2/t_1}$ and $\abs{t_1}/t$ along both lines are plotted as a function of $\Omega$ in Fig. \ref{HCB_regime}(b). In both Fig. \ref{HCB_regime}(a) and Fig. \ref{HCB_regime}(b), the blue triangle and the yellow triangle are placed at the points where the SF-BO transition takes place along the $\delta = 0.15 t$ trajectory. The red square indicates the position of the integrable MG point at $\delta = 0$. We observe that a return to superfluidity after the SF-BO transition is expected for small and nonzero Raman detuning, which can be in principle explored in realistic experimental setups. In Fig. \ref{HCB_regime}(c) the extrapolated energy gap $G_{\infty}$ and the bond-order parameter $O_{BO}$ (see \eqref{BOparam}) for a chain of size $L=100$ are shown as a function of $\abs{t_2/t_1}$ across the corresponding trajectories. 

\begin{figure}[b]
\includegraphics[width=0.75\linewidth]{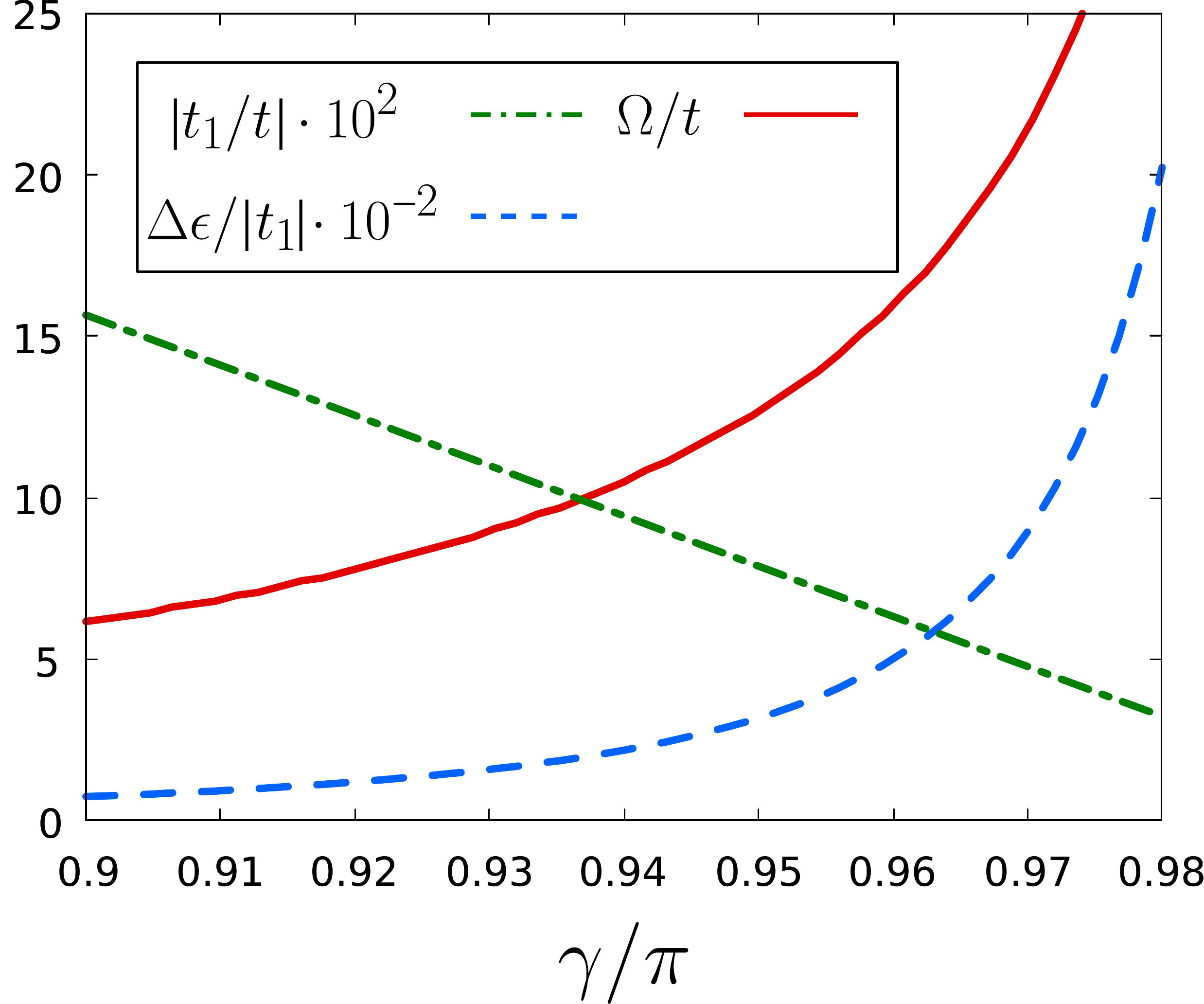}
\caption{(Color online) \textbf{Energy scales at the MG point:} Set of values of $\Omega$, $|t_1/t|$ and $\Delta \epsilon /|t_1|$ as a function of $\gamma$ compatible with the MG point with $|t_2/t_1|=0.5$ and $\Phi = \pi$. The dashed blue curve indicates the scale separation between the characteristic scale of the effective ladder $|t_1|$ and the gap to the higher band in the original ladder $\Delta \epsilon$. The HCB and the lowest-band approximations taken simultaneously require  $|t_1|\ll U \ll \Delta_\epsilon$. Thus, on the one hand the dashed blue curve should be indicatively above one. On the other hand,  too large values of $\Delta \epsilon \sim 2\Omega$, challenge the tight-binding condition. See the main text for further details.}\label{HCB_gamma} 
\end{figure}

For the trajectories plotted in Fig. \ref{HCB_regime}, we have fixed the Raman detuning $\delta$ and momentum $k_R$ and scanned with the Raman intensity $\Omega$ as a single parameter. This is experimentally convenient as the whole set of points can be covered by varying only one dynamically adjustable parameter. In turn, as displayed in Fig. \ref{HCB_regime}(b), $\abs{t_1}/t$ has a very soft dependency on $\Omega$ at a given value of $\gamma$ (the bandwidth decreases very slowly at large $\Omega$). This allows to keep the same characteristic time scale $\propto 1/\abs{t_1}$ along the whole trajectory at fixed $\gamma$. Notice, although, that the phase diagram in Fig. \ref{HCB_regime}(a) is represented as a function of only the two parameters $\Phi$ and $\abs{t_2/t_1}$. This means that each point corresponds in fact to a collection of points in parameter space $(\Omega, \gamma,\delta)$. For the model developed here, the set of suitable values for $\Omega$ is lower-bounded by the HCB approximation and upper-bounded by the tight-binding approximation. The accuracy of the former increases with larger $\Omega$ ($\Delta \epsilon /\abs{t_1} \sim 2 \Omega /\abs{t_1}$), while the opposite is true for the latter. Thus, for any given point in the diagram, it can be convenient to adjust $\gamma$ to allocate $\Omega$ within a suitable range, as shown in Fig. \ref{HCB_gamma}. 
 
Let us briefly comment on the expected experimental uncertainties in determining the parameters $(\Omega, \gamma,\delta)$ and thus the phase diagram in Fig. \ref{HCB_regime}. In experiments, the laser wavelengths and angles of incidence, and thus $\gamma \propto \lambda_L \cos(\theta_R)/(\lambda_R \cos(\theta_L))$, can be adjusted essentially with arbitrary precision. Furthermore, expected minor uncertainty in the Rabi coupling ($\delta\Omega\lesssim\pm 1 t \sim 0.03 E_R$ for a real-space tunnelling $t\sim 10^2 Hz$ are currently achievable \cite{Campbell-2016}) does not hinter the location of the BO phase and reflect in minor uncertainties in the location of the phase boundary.  The more sensitive parameter is thus the detuning $\delta$, whose stability depends on the bias magnetic field responsible for the Zeeman split between the two internal states employed. Given that the BO phase predicted here is only realized very close to $\delta=0$, uncertainties in $\delta$ should not exceed $\pm 0.1t$. The sensitivity to magnetic fluctuations can be downplayed for instance by employing rf-dressed states \cite{Campbell-2016-njp, Trypogeorgos-2018-pra,Anderson-2018-pra}.

Finally, we would like to stress that the HCB regime studied here is assumed within the triangular ladder mapping: we do not require that $U\gg t$, since $t_1, t_2$ can be much smaller than $t$. This supposes a departure from the treatment of the two-leg ladder with flux done in \cite{Piraud-2015, Piraud-2016}, where they assume the onsite interaction energy per pair to be much larger than both the interleg and intraleg tunneling ratios to study the HCB limit. Moreover, the authors consider onsite interactions in each site of the rectangular ladder, unlike the case with a synthetic second dimension that we consider. At the single-particle level both systems are equivalent. However, in the case of two real dimensions the interacting energy per pair of sites is halved with respect to the synthetic dimension scenario presented here. This makes the HCB approximation within the effective triangular ladder more restrictive in the realisation there discussed, as larger intraleg-interleg tunneling ratios ($\Omega/t$) would be required to achieve the same regime, further threatening the tight-binding approximation. Thus, a ladder with synthetic dimensions appears to be favorable in order to realize the physics we describe in this section.

\subsection{Properties of the bond-ordered phase}

The BO insulating phase is better understood by looking at the properties of the state at the MG point, at $\Phi = \pi$ and $t_2 = -0.5 t_1 < 0$, where the system is solvable \cite{Majumdar_1970}. In the thermodynamic limit ($L\rightarrow \infty$), the ground state is degenerate at this point. The two ground states $\ket{\psi_{\text{e,o}}}$ are given by a product state of triplet spin states defined on pairs of consecutive sites 
\begin{equation}\label{MGeigenstates}
\ket{\psi_{\text{e,o}}} = \bigotimes_{j\in \text{even} / \text{odd}}\frac{\ket{\uparrow_{j}\downarrow_{j+1}}+ \ket{\downarrow_{j}\uparrow_{j+1}}}{\sqrt{2}}.
\end{equation}
At finite chain sizes, the degeneracy is broken when the chain has an even number of sites. Still, the bond order parameter $O_\mathrm{BO}$ is exactly $0.5$ at the MG point, regardless of the length of the chain.

The bond-ordered phase can be regarded as a valence bond crystal \cite{Lacroix-2011} with dimerized triplet states instead of singlets. In the dressed-atom basis, such states correspond to having an atom per pair of sites, oscillating between each site within the pair (the dressed-atom basis is essentially determined in terms of the original spin states by the eigenstates of the single-particle Hamiltonian \eqref{hamqspin12}). 

It is well known that dimerization transitions take place in spin chains with long-range interactions with sufficiently strong frustrated next-nearest-neighbor interactions \cite{Majumdar-1969}. In frustrated spin chains, the dimerization transition is driven by a perturbation which becomes marginally relevant in the bond-ordered phase, with an initially exponentially small energy gap \cite{Affleck-1987-prb}. Interestingly this behavior is reproduced by Hamiltonian \eqref{hardcoreham} for $\Phi\lesssim \pi$, as illustrated in Fig. \ref{HCB_regime}(c).

The HCB Hamiltonian \eqref{hardcoreham} for $\Phi = \pi$ describes a particular case of long-range interacting spin chain
\begin{align}\label{hardcoreham_interacting}
H = & \sum_j \left( -J_x S_{j}^{x}S_{j+1}^{x} - J_y S_{j}^{y}S_{j+1}^{y}\right) \cr
& + \sum_j \left(K_x S_{j}^{x}S_{j+2}^{x} + K_y S_{j}^{y}S_{j+2}^{y}\right) .
\end{align}
with $J_x = J_y = t_1$ and $K_x = K_y = t_2$. The above Hamiltonian has been studied in detail in \cite{Jiang-2019-prb} as a 1D analogue of the deconfined quantum criticality \cite{Senthil-2004-Science} (for an extension to power-law decay couplings see \cite{yang-2020-arxiv}). 

Note that the BO phase can be determined by measuring the bond-order parameter \eqref{BOparam} in the original two-leg square ladder.  The measurement can be performed by applying similar techniques as in \cite{Sebby-2006-pra, folling-2007-nature, Greif-2013-science,Atala-14}. Here, in order to project the state on the bonds and detect the order by measuring the oscillation of the population under a tilting, in addition to a superlattice that disconnects even and odd plaquettes, one has also to rotate the dressed states into the original spin states. The same method can be also employed to measure the current (see below). Alternatively, instead of directly measuring the current, one could study the response of the system to a perturbation and measure  it spectroscopically, as suggested in  \cite{Loida-2017}.
\begin{figure}[t]
\includegraphics[width=1.0\linewidth]{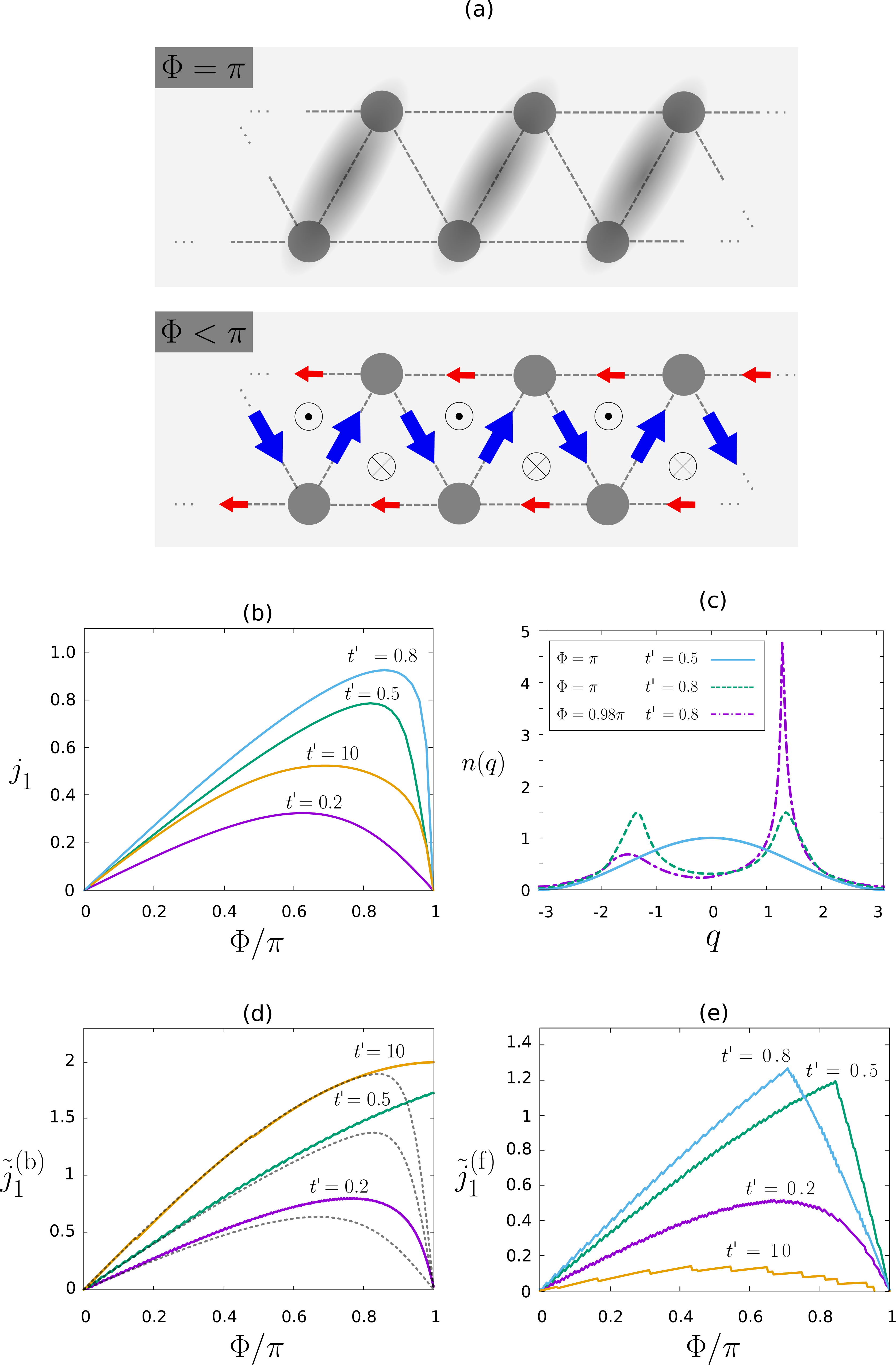}
\caption{(Color online) \textbf{Currents distribution in the effective ladder:} (a) Sketch of the current patterns in the ground state of the HCB hamiltonian \eqref{hardcoreham}. At frustration, the currents are suppressed (top). The shadowed links represent the dimerized pairs. For $\Phi<\pi$, the ground state presents a uniform zigzag current $j_1$ (blue arrows) and arm current $j_2 = -j_1/2$ (red arrows). (b) Current per particle $j_1$ as a function of $\Phi$ for different values of $t'= |t_2/t_1|$ and for $L=200$. (c) Momentum distribution $n(k)$ of the ground state for $t'=0.5$ and $t'=0.8$. For $\Phi=\pi$, a commensurate-to-incommensurate transition occurs at $t'=0.5$, characterized by the onset of two peaks. (d-e) Equivalent currents for a non-interacting Bose (d) and Fermi (e) gas at zero temperature. The dashed lines in (d) represent the departure from the zero temperature solutions at finite temperature. In (e) the currents are evaluated at quarter filling. In both cases, the discontinuities in the currents are due to the finite size of the chains.}\label{triangular_ladder_currents} 
\end{figure}

\subsection{Current distribution in the triangular ladder}

Now we look at the current distribution in the ground state in the HCB limit, and compare it to the single-particle scenario discussed in Sec. \ref{sec-spin1/2}. In the HCB mapping, the spin current from site $i$ to site $i+l$ in the lattice is given by
\begin{equation}
\mean{j_{i,i+l}} = -2\abs{t_{l}} \textrm{Im}\left(\enum{i\phi_{l}} \mean{S_i^{-}S_{i+l}^{+}}\right).   
\end{equation}

The system presents a ``zigzag'' rung current $j_{i,i+1}$ in one direction and an arm current $j_{i,i+2}$ in the opposite direction, as schematically represented in Fig. \ref{triangular_ladder_currents}(a). For the finite-size chains, we define the currents per particle as $j_{l} = \frac{2}{L}\sum_i j_{i,i+l}$. Across the whole phase diagram the density is uniform and the net current between consecutive dimerized pairs of sites is zero, which implies that $j_{i,i+1} = -2j_{i,i+2}$, for any $i$.

The current $j_{1}$ is represented in Fig. \ref{triangular_ladder_currents}(b) as a function of the flux $\Phi$. For comparison, the analogous single-particle current $\tilde{j}^{(b)}_1 = \partial E_g/\partial \phi_1$ is represented in Fig.  \ref{triangular_ladder_currents}(d). Here, $E_g = \mean{H_\mathrm{n.i.}}_g$ is the ground state energy per particle for the single-particle Hamiltonian \eqref{niham_ladder}. At zero temperature, this energy corresponds to the minimum of the band \eqref{effectiveband}. For non-interacting fermions, the currents are given by $j^{(f)}_1 = \frac{2t}{L} \sum_q n_q \cos(q+\Phi/2)$ and $j^{(f)}_2 = -\frac{2t}{L} \sum_q n_q \cos(2q)$, where $n_q$ is the occupancy of the band mode $q$. At quarter filling, $j^{(f)}_1$ is represented in Fig. \ref{triangular_ladder_currents}(e). The two-minima regime in the dispersion band can be distinguish by the presence of a discontinuity in the derivative of the current $\partial j^{(f)}_1(\Phi)/\partial \Phi$. 
As expected, when $t_2$ is small, the HCB system behaves similarly as a non-interacting fermion gas (superfluid phase and similar current distribution). As $t_2$ is made larger, the system departs from the free fermion picture, and correlated terms become increasingly dominant. Fermions and hard-core bosons behave similarly regarding their local density and currents. However, they differ considerably in momentum space, with bosons exhibiting a peaked distribution. 
In Fig. \ref{triangular_ladder_currents}(c), we plot the momentum distribution $n(q) = \frac{1}{L}\sum_{ij}\enum{iq(i-j)} \mean{S_i^{+}S_j^{-}}$. At $t_2 = 0.5 t_1$ the system undergoes a commensurate-to-incommensurate transition, characterised by the onset of two peaks for larger values of $t_2$. Similarly as with free fermions, the equal population of the two peaks at frustration prevents the system from having nonzero current densities. When the system moves away from frustration, i.e., $\Phi\neq (2k+1)\pi$ the current density rapidly increases. 

In the non-interacting limit, by contrast, a superposition of the two plane waves in the two-degenerate-minima regime ($\abs{t_2} > 0.25 \abs{t_1}$) cannot be stabilized at zero temperature, and the system spontaneously occupies one of the two minima, which yield non-vanishing currents even at frustration. Contrarily, interactions protect against the spontaneous occupation of a single peak in the HCB system. This picture is partly recovered for free bosons at finite temperature (see Fig.  \ref{triangular_ladder_currents}(d)). However, notice how for both HCB and free fermions the current decreases with $t_2$ after reaching its maximum value, while the single particle current converges fast to $j^{(b)}_1 \overset{t_2\rightarrow \infty}{\longrightarrow}  2t \sin(\Phi/2)$.

Contrary to both the free boson and fermion cases, the behavior of the current in the HCB limit does not capture the commensurate to incommensurate transition. In addition, correlations appear to have a minor role on the current and density distribution in the ground state. 


\section{Discussion and outlook}\label{sec-outlook}

In this work we have shown that Bose gases with Raman-induced artificial SOC in 1D lattices can be employed to explore the physics of two-leg triangular ladder configurations. By controlling the strength, the detuning, and the Raman momentum kick $k_R$ (i.e., the angle of incidence of the Raman dressing beams), the different parameters of the effective triangular ladder can be widely adjusted, including a staggered flux that can range from $\pi$ at the Raman resonance to zero. The mapping is obtained in the spin-$1/2$ system following a lowest-band truncation of the Hamiltonian. While in the original semisynthetic square ladder the interactions have an intrinsic nonlocal character, in the truncated ladder they can be made onsite, even in the case of asymmetric interactions between the spin states. Exploiting the large separation between the bandwidth of the lowest band and the energy gap to the higher band, we have shown that both the weakly and the strongly interacting regimes of the triangular ladder can be covered within the mapping.

Following a hard-core boson treatment of the effective ladder, we have studied the properties of the resulting spin chain. At half-filling, the fully frustrated ladder realizes an antiferromagnetic $XX$-spin model, which displays a superfluid phase and a bond-order gapped phase separated by a BKT-type phase transition. We show that the bond-order phase is accessible in the SOC system and can also be stabilized for fluxes close but different from $\pi$. 
Notably, the phase arises for moderate interactions in the semisynthetic square ladder (the hard-core boson regime is considered within the effective triangular ladder). This supposes a departure from the conventionally explored strongly interacting regime in the two-leg ladder with flux. To the best of our knowledge, such dimerized phase has not been described previously for SOC Bose gases in the lattice.

Our study opens interesting perspectives. On the one hand, it shows that semisynthetic SOC coupled lattices provide an alternative, practical, and scalable platform for the experimental realization of frustrated spin chains/triangular ladders with staggered fluxes, beyond trapped ions (where the flux can be engineered by a periodic driving as suggested in \cite{Grass-15-pra,Grass-18-pra}), Rydberg atoms (where the flux can be engineered by exploiting the angular dependence of the Rydberg interaction between $p$-wave Rydberg state, as  very recently experimentally demonstrated in \cite{Lienhard-2020}), and cold atoms trapped along photonic crystals waveguides (as proposed in \cite{Hung-2016-PNAS}, where the interactions are mediated by guided virtual photons in a photonic band gap). On the other hand, our study poses intriguing theoretical questions, for instance, about the fate of the deconfined critical phase transition for fluxes $\Phi\neq \pi$, and about the nature and the interpretation of BO phase from the fractional quantum Hall perspective.

To this end, it can be extremely helpful to repeat the derivation of the effective Hamiltonian for the triangular ladder from the square flux ladder using bosonization and/or renormalization group approaches \cite{Giamarchi-2003-book,Cazalilla-2011-rpm}, see for instance \cite{Citro-cm-2020spectral} and \cite{Tirrito-prb-2019} for recent applications of these techniques to quasi-1D bosonic and fermionic systems, respectively.

\acknowledgements
\small
J.Cabedo, A.C., V.A. and J.M. acknowledge support from the Ministerio de Economía y Competividad MINECO (Contract No. FIS2017-86530-P), from the European Union Regional Development Fund within the ERDF Operational Program of Catalunya (project QUASICAT/QuantumCat), and from Generalitat de Catalunya (Contract No. SGR2017-1646). J.Claramunt acknowledges partial support from DGI-MINECO-FEDER (Grant MTM2017-83487-P) and the research funding Brazilian agency CAPES. A.C. acknowledges support from the UAB Talent Research program.
\normalsize

\section*{Author contribution statement}

J. Cabedo and A.C. designed the project and wrote the draft. J.Cabedo and J.Claramunt made the calculations. J.Claramunt, V.A., J.M. participated in the editing and revision of the main text.


%

\end{document}